\documentclass[prb]{revtex4} \usepackage{graphicx}
\begin{document}
\title{Density of states of a binary Lennard-Jones Glass}
\author{Roland Faller} \affiliation{ Department of Chemical
Engineering \& Materials Science, University of California-Davis,
Davis, CA 95616} \author{Juan J. \surname{de Pablo}}
\affiliation{Department of Chemical Engineering, University of
Wisconsin-Madison, Madison, WI 53706}
\begin{abstract}
We calculate the density of states of a binary Lennard-Jones glass
using a recently proposed Monte Carlo algorithm. Unlike
traditional molecular simulation approaches, the algorithm samples
distinct configurations according to self-consistent estimates of
the density of states, thereby giving rise to uniform
internal-energy histograms. The method is applied to simulate the
equilibrium, low-temperature thermodynamic properties of a widely
studied glass former consisting of a binary mixture of
Lennard-Jones particles. We show how a density-of-states algorithm
can be combined with particle identity swaps and configurational
bias techniques to study that system. Results are presented for
the energy and entropy below the mode coupling temperature.
\end{abstract}
\maketitle
\section{Introduction}
The transition from a liquid to an amorphous solid that sometimes
occurs upon cooling remains one of the largely unresolved problems
of statistical physics~\cite{ediger96,debenedetti01}. At the
experimental level, the so-called glass transition is generally
associated with a sharp increase in the characteristic relaxation
times of the system, and a concomitant departure of laboratory
measurements from equilibrium. At the theoretical level, it has
been proposed that the transition from a liquid to a glassy state
is triggered by an underlying thermodynamic (equilibrium)
transition~\cite{mezard99}; in that view, an ``ideal'' glass
transition is believed to occur at the so-called Kauzmann
temperature, $T_K$. At $T_K$, it is proposed that only one
minimum-energy basin of attraction is accessible to the system.
One of the first arguments of this type is due to Gibbs and
diMarzio~\cite{gibbs58}, but more recent studies using replica
methods have yielded evidence in support of such a transition in
Lennard-Jones glass formers~\cite{mezard99,coluzzi00a,grigera01}.
These observations have been called into question by experimental
data and recent results of simulations of polydisperse hard-core
disks, which have failed to detect any evidence of a thermodynamic
transition up to extremely high packing fractions~\cite{santen00}.
One of the questions that arises is therefore whether the
discrepancies between the reported simulated behavior of hard-disk
and soft-sphere systems is due to fundamental differences in the
models, or whether they are a consequence of inappropriate
sampling at low temperatures and high densities.

Different, alternative theoretical considerations have attempted
to establish a connection between glass transition phenomena and
the rapid increase in relaxation times that arises in the vicinity
of a theoretical critical temperature (the so-called
``mode-coupling'' temperature, $T_{MCT}$), thereby giving rise to
a ``kinetic'' or ``dynamic'' transition~\cite{goetze92}. In recent
years, both viewpoints have received some support from molecular
simulations. Many of these simulations have been conducted in the
context of models introduced by Stillinger and Weber and by Kob
and Andersen ~\cite{kob95a}; such models have been employed in a
number of studies that have helped shape our current views about
the glass
transition~\cite{sastry98,sciortino99,donati99,coluzzi00a,coluzzi00b,yamamoto00}.
The particular model considered here consists of a binary mixture
of Lennard-Jones particles, with composition 80\% $A$ and 20\%
$B$. A total of 250 particles is employed in our calculations. The
interaction parameters between particles of species $A$ and $B$
are $\epsilon_{AA}=1.0$ and $\sigma_{AA}=1.0$, $\epsilon_{BB}=0.5$
and $\sigma_{BB}=0.88$, and $\epsilon_{AB}=1.5$ and
$\sigma_{AB}=0.8$. The density is $1.204\sigma_{AA}^{-3}$.
Recently, a crystal structure at extremely low energies has been
reported for a variant of this system~\cite{middleton01}.

High-precision data are available for the thermodynamic properties
of this model at intermediate to high
temperatures~\cite{kob95a,yamamoto00,coluzzi00b}. A series of
careful simulations have placed the mode coupling temperature at
$T_{MCT}=0.435$ and the Kauzmann temperature somewhere in the
range $T_K=0.26-0.31$~\cite{yamamoto00,sciortino99,coluzzi00b}.
Note, however, that literature studies have generally avoided
direct simulations below $T_{MCT}$; available estimates of $T_K$
have been produced after making several assumptions regarding the
potential energy landscape and by {\it extrapolation} (to low
temperatures) of liquid-state data generated at higher
temperatures (above $T_{MCT}$). An exception is provided by a
recent report for a related model~\cite{grigera01}, where
simulations of small systems, directly at low temperatures,
suggest that an anomaly in the heat capacity $c_v$ arises at
$T_K$; $c_v$ is reported to increase with decreasing temperature,
and to exhibit a sharp drop at $T_K$. The drop becomes more
pronounced as the system size is increased.

Simulations near a glass transition are notoriously difficult, and
their results must be considered with caution.  On the one hand,
the relevant time scales below $T_{MCT}$ are too long to be
sampled by conventional molecular dynamics simulations.  Monte
Carlo techniques, on the other hand, have been used only rarely to
simulate glass formers; furthermore, it has been difficult to
establish to what extent available studies have succeeded in
sampling relevant regions of phase space, particularly at low
temperatures and elevated densities. In this work, we use a novel
Monte Carlo sampling technique to arrive at {\it direct} estimates
of the thermodynamic properties of a model glass former down to
temperatures well below $T_{MCT}$.
\section{Simulation Methods}
Recently, Wang and Landau have proposed an iterative method to
estimate the density of states of a Potts lattice system from a
Monte Carlo simulation~\cite{wang01a,wang01b}. The random-walk
algorithm is based on the idea of entropic sampling, with a
self-consistent update of the density of states. It has proven to
be remarkably efficient for lattice systems, simple
liquids~\cite{yan02}, proteins~\cite{rathore02,rathore03}, and
liquid crystals~\cite{kim02}; it is tempting to apply it in the
context of a glass-forming liquid. In this contribution we combine
it with biased sampling techniques, and we use it to generate {\it
direct} estimates of the density of states of the glass-former
described above.

In a conventional canonical-ensemble simulation, different states
of the system are visited with probability $\Omega (E)
e^{-E/k_BT}$, where $\Omega (E)$ is the density of states (or
degeneracy) of the system, $k_B$ is Boltzmann's constant, and $T$
is the temperature. In contrast, in the random-walk scheme adopted
here, the density of states $\Omega (E)$ is estimated directly by
producing a uniform, or ``flat'' histogram of energies, i.e. by
coercing the system to visit all energy states with equal
probability. In this study we have chosen to maintain a constant
density and constant number of particles; extensions to other
physical ensembles and to expanded ensembles have also been
pursued recently~\cite{yan02,kim02,calvo02}. Trial moves are
generated by means of simple translations of the particles and by
identity interchanges~\cite{grigera01}. The acceptance of such
interchanges is enhanced using configurational bias. The resulting
trial configurations are accepted with probability~\cite{jain02}
\begin{equation}
  p=\text{min}\Big\{1,\frac{W^R_{new}\Omega^{\prime}(E_{old})}
  {W^R_{old}\Omega^{\prime}(E_{new})}\Big\}
\end{equation}
where the prime indicates that this is a transient, momentary
``best estimate'' of the density of states. Biased moves are
performed according to a Rosenbluth type algorithm; $W^R$ is the
Rosenbluth weight of the corresponding state. In the original
version by Wang and Landau $W^R$ was set to unity for all states.
The configurational bias identity swap consists of the following
steps: First a pair on unlike particles is chosen at random. After
direct interchange of their positions, the smaller particle $B$
will fit in the cavity formerly occupied by the larger particle
$A$. The opposite is only seldom true, thereby leading to
negligible acceptance rates at low energies and high densities. In
order to enhance the acceptance rate, $N_{CCB}=12-100$ trial
positions are explored for the $B$ particle around the position
formerly occupied by the $A$ particle. We then apply
configurational bias ideas to these $N_{CCB}$ positions. There are
now two possible ways to calculate the Rosenbluth factors. In the
first of these, the energy of the states can be used directly (as
is done in standard configurational bias Monte
Carlo)~\cite{depablo92,frenkel96}. To this end, a fictitious
temperature $T_F$ is introduced for calculation of the Rosenbluth
factor $W^R_i$ for state $i$
\begin{equation}
W^R_i=\frac{\exp(-\beta_FE_i)}{\sum_j\exp(-\beta_FE_j)}.
\end{equation}
Note that this fictitious temperature is not the temperature of
the system, although in conventional configurational bias
simulations it is set to the system temperature. Alternatively,
one can avoid using a fictitious temperature by calculating a set
of $W_R$ using the density of states itself as a bias.
\begin{equation}
W^R_i=\frac{\Omega_i}{\sum_j\Omega_j}.
\end{equation}
Both variants (using $T_F=0.5-1$) behave similarly. The acceptance
rate is extremely small (it drops to less than $10^{-6}$ as the
effective temperature approaches $T_g$), but it is sufficient to
perform simulations over extended amounts of computer time. The
effective temperature $T_{\text{eff}}$ of a state with energy $E$
is defined as the temperature where $\langle
E_{pot}\rangle_{T_{\text{eff}}}=E$.

The density of states is not known \'a priori; it is initially set to
unity throughout the entire energy range. The calculations begin by
defining an energy range in which to determine $\Omega(E)$. Whenever
an energy state $E$ is visited, the density of states
$\Omega^{\prime}(E)$ corresponding to that energy is multiplied by a
constant $f$, i.e.  $\Omega^{\prime}(E) \rightarrow f
\Omega^{\prime}(E)$. Since the density of states varies over many
orders of magnitude, it is convenient to work with its logarithm,
which corresponds to the entropy as a function of energy $S(E)=-k_B
\ln \Omega(E)$. The entropy is updated by adding a constant $\ln f$.

A histogram of energies $h(E)$ is also constructed and updated after
every trial move. The density of states is updated continuously
throughout the simulation, until the recorded energy histogram $h(E)$
is sufficiently flat. Note that in actual practice it is not possible
to generate a perfectly flat histogram; in this work, flatness is
considered to be attained if the minimum of $h(E)$ is at least 0.9
times the average value $\overline{h(E)}$. Having reached a
``flat-histogram'' condition, the simulation sequence is repeated: the
energy histogram is erased, and the new value of $\ln f$ is set to
half the ``old'' value. Note that this choice is arbitrary, and any
monotonically decreasing function should work. The initial value of
$\ln f$ was set to unity, and the final value was $\ln f = 10^{-5}$,
which corresponds to 18 iterations. The factor $\ln f$ controls the
convergence of $\Omega^{\prime}(E)$ to the true value, $\Omega(E)$; as
$\ln f$ decreases (i.e. as the simulation proceeds) the calculations
per iteration become increasingly long.

\begin{figure}
  \includegraphics[width=0.45\linewidth]{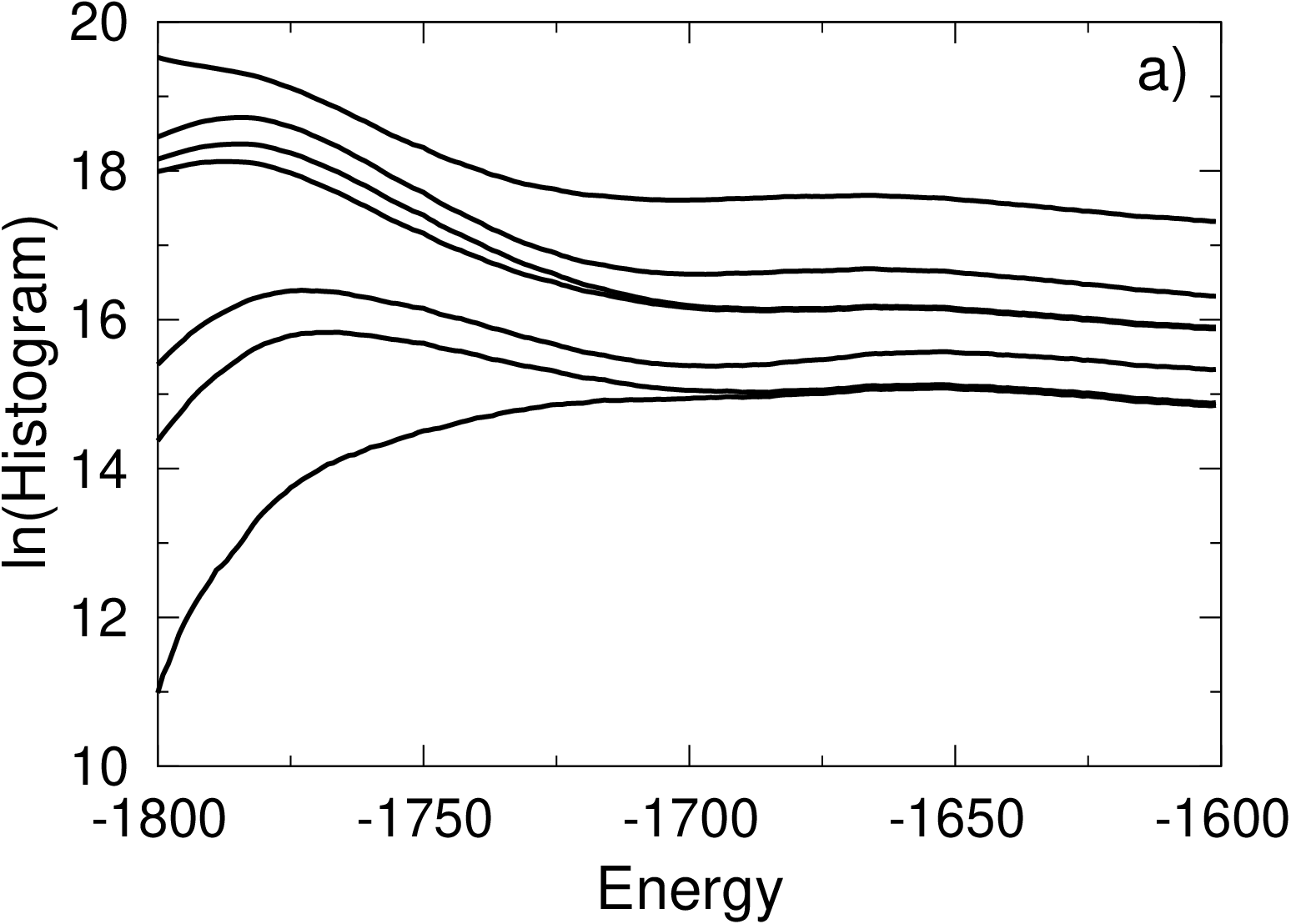}
  \includegraphics[width=0.45\linewidth]{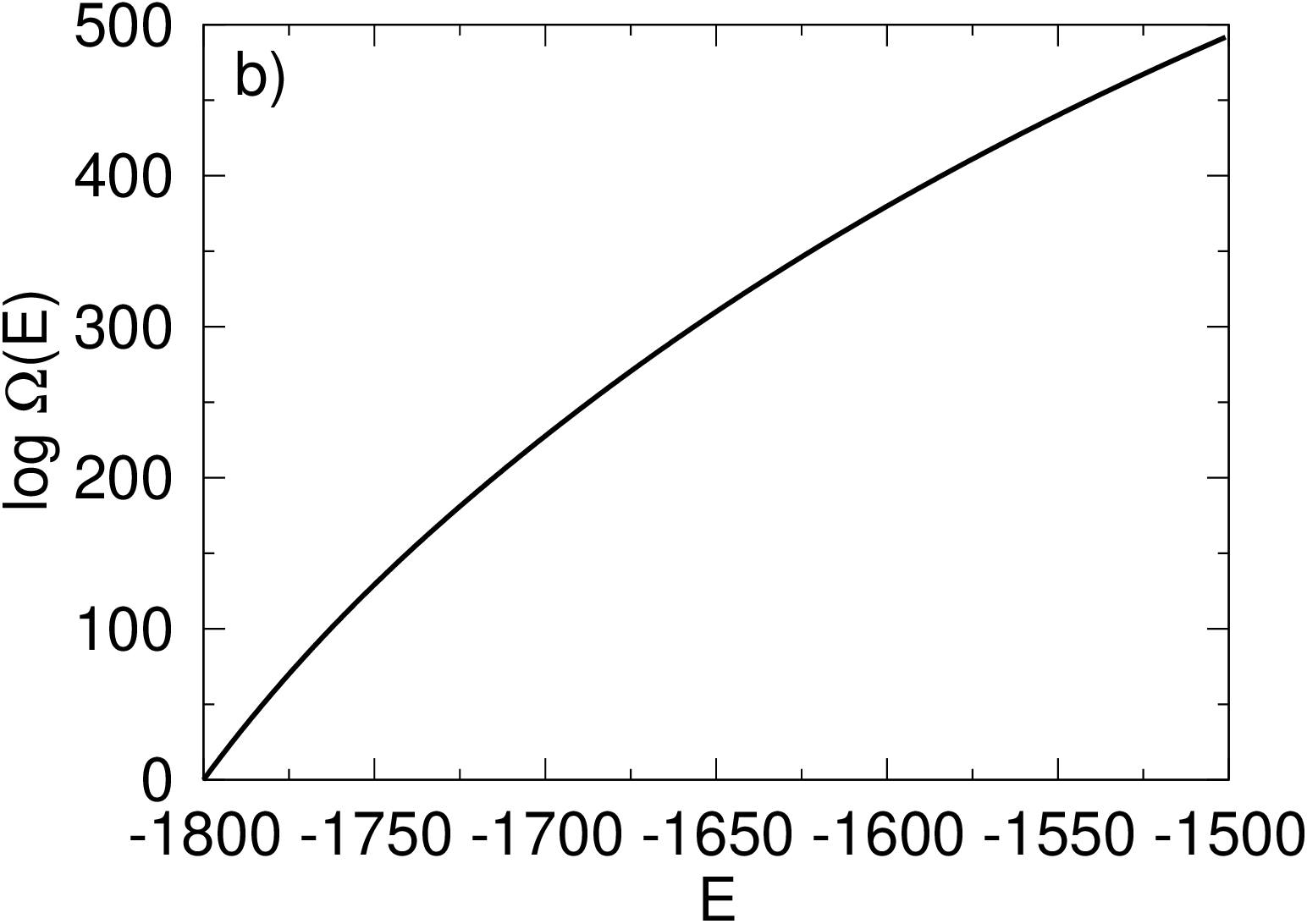}
  \caption{ a) Logarithm of the energy histogram in the final run
  without updating the weights in different stages of the
  simulation. The curves are parallel to each other. b) Logarithm of
  the density of states over the decisive energy range [-1800,-1500].}
  \label{fig:dos}
\end{figure}

To improve efficiency, it is useful to conduct multiple
simulations in overlapping energy ranges. These ranges must be
relatively narrow; otherwise, given the rapidly varying nature of
$\Omega(E)$, the calculations can be prohibitively long. The
energy ranges employed here correspond to
$E_{\text{max}}-E_{\text{min}}=200-400$, or
$(E/N)_{\text{max}}-(E/N)_{\text{min}}=1.0-2.0$ per particle. In
order to generate $\Omega(E)$ over a wide energy range,
neighboring energy windows were constructed in such a way as to
overlap by half their width; every region of the energy axis is
covered by at least two independent simulations. Two sets of
independent simulations with energy windows of 200 or 400 units
wide were pursued here. At low energies, convergence was only
possible with relatively narrow energy ranges. The lowest energy
range employed here starts at $E=-1860$ ($E/N=-7.44$), which is
slightly above the range of estimated intrinsic energies for this
system~\cite{sciortino99} and well above the crystal
energies~\cite{middleton01}. After the density of states converged
to within a certain accuracy the update of $\Omega (E)$ was
stopped. A simple multicanonical run was then employed, where the
inverse of the DOS was used as the weighting function. After a
long run (25 million steps) the density of states was corrected by
adding the logarithm of the final histogram. This was necessary as
we did not run the simulations to update factors of
$f=\exp(10^{-8})$ as in the original work by Wang and
Landau~\cite{wang01a,wang01b}. It was verified during the course
of the simulation that the logarithmic energy histograms kept
increasing homogeneously (Figure~\ref{fig:dos}a).  This was
monitored by observing that the logarithms of the histograms taken
at different points in time during the final run are filled
homogeneously. In a theoretically ideal situation, the different
(logarithmic) histograms would be parallel to each other. However,
the stochastic nature of the simulation leads to departures from
parallel behavior. Nonetheless, we see that the system never gets
``trapped'' in a certain energy region without visiting the
others. This ensures that the system readily moves back and forth
between high and low energy regions of phase space, which is
analogous to moving through temperatures in other types of
simulations. In the case of continuous degrees of freedom this
additional monitoring is important.

\begin{figure}
  \includegraphics*[width=0.47\linewidth]{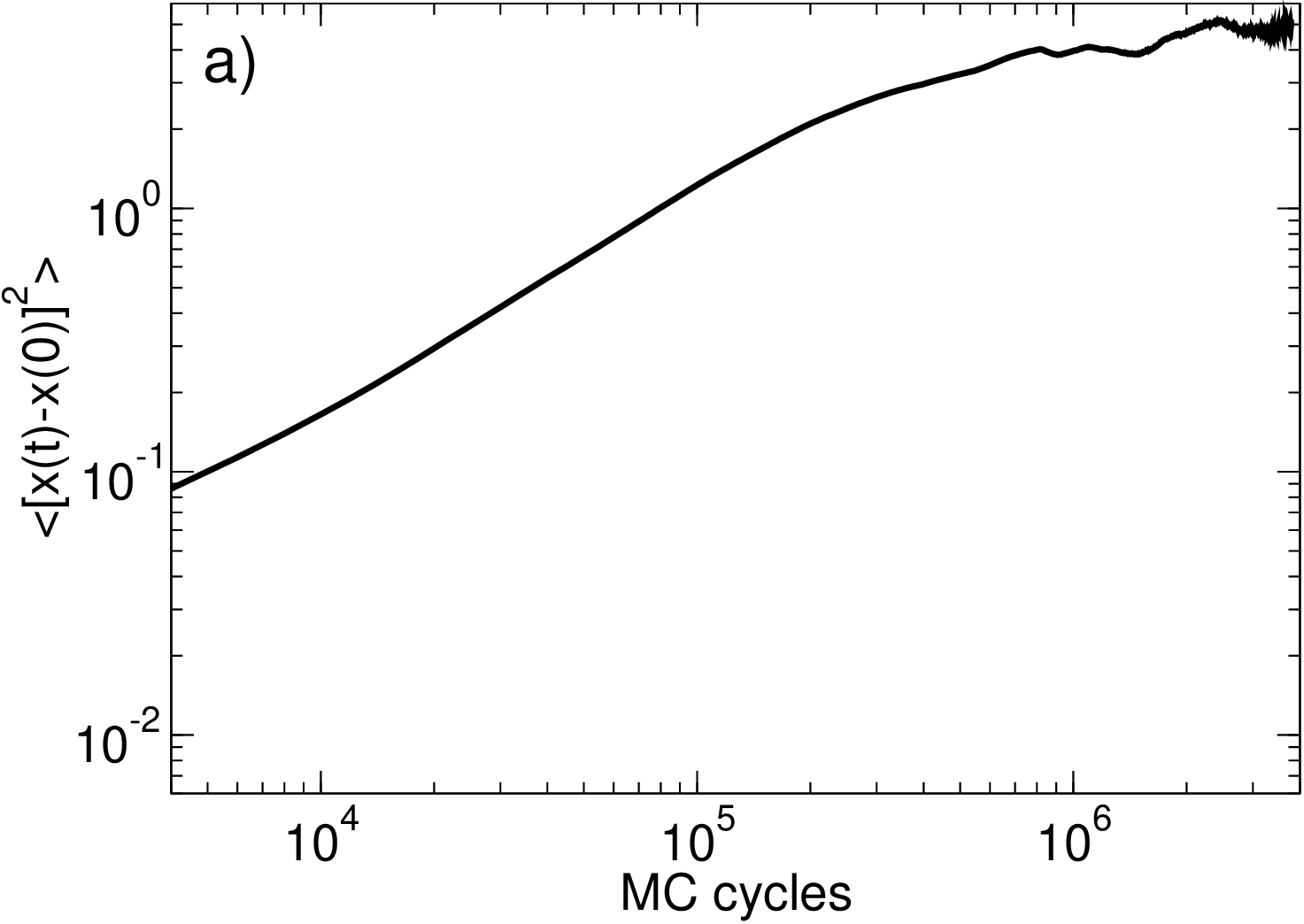}\hspace{0.5cm}
  \includegraphics*[width=0.47\linewidth]{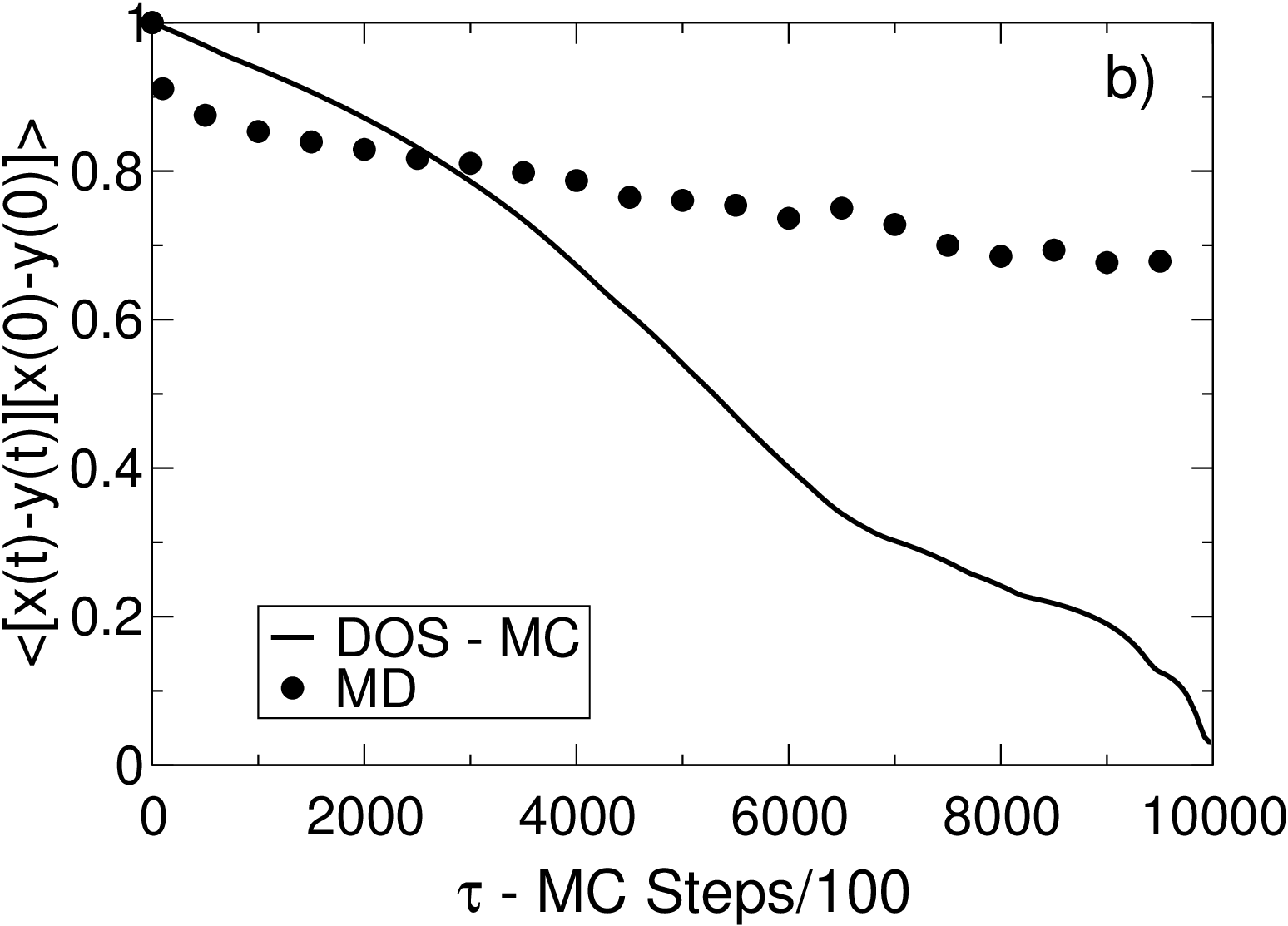}
  \caption{a) Average mean-squared displacement in the final
  multicanonical run (see text). The length of the simulation box is
  5.96. The mean-squared displacement is calculated in the original
  box so it apparently levels off approaching the box size. b)
  Position autocorrelation function from Monte Carlo simulations and
  from molecular dynamics at $T=0.45$.}
  \label{fig:msd}
\end{figure}

In order to provide an assessment of the correlation time of the
random-walk method, the mean squared displacement of the particles
was also measured. It is observed that in the final multicanonical
run approximately 2 million cycles are necessary for the
mean-squared displacement to be comparable to the box length (in
one MC cycle each particle is moved once). The simulations
presented here are at least 10 times that length (cf.
Figure~\ref{fig:msd}). A positional autocorrelation function can
be defined as
\begin{equation}
C_{pos}(t)=\langle(\vec{x}_i(t_0+t)-\vec{x}_j(t_0+t))
(\vec{x}_i(t_0)-\vec{x}_j(t_0))\rangle_{i,j,t_0}.
\end{equation}
This measure of relaxation is more stringent than the mean-square
displacement as it eliminates the possibility that blocks of
particles might be moving together without too much mutual
rearrangement. This function decays to zero within the simulation
lengths considered here. Figure~\ref{fig:msd} also compares this
function to that obtained from molecular dynamics runs at
$T=0.45$. To the best of our knowledge, the longest runs reported
in the literature~\cite{yamamoto00} for the system considered here have lasted
$10^5\tau$ ($\tau$ is the usual dimensionless
Lennard-Jones time, which typically corresponds to 100 timesteps).
In $10^5\tau$, this function decays to about 70\% of its initial
value. The scale employed in Figure~\ref{fig:msd} assumes that the
computational requirements for one MD timestep are comparable to
those of one MC cycle.

In regions of overlap, the density of states corresponding to each window can
only differ by a constant, which depends on the (arbitrary) number of
histogram entries. The density of states over the entire energy range of
interest is constructed by shifting local estimates of $\ln \Omega(E)$
(corresponding to individual windows) until they coincide, in the middle of the
overlap region.

The global density of states is therefore known to within a constant. Since
internal energies are known exactly, the excess free energy of the system can
be calculated as a function of $T$ according to
\begin{equation}
  \beta \langle F(T)^{\text{ex}}\rangle=-\ln \;\sum_E \Omega(E)\; e^{-\beta E},
\end{equation}
where the brackets denote an ensemble average and where $\beta=1/k_BT$.
Similarly, the average internal energy of the system is given by
\begin{equation}
 \beta \langle E(T)^{\rm ex}\rangle =\frac{\sum_E E \;\Omega(E)\; e^{-\beta E}}
{\sum_E \Omega(E) \; e^{-\beta E}} ,
\end{equation}
and the entropy can simply be determined from $\langle
S\rangle=(\langle U\rangle-\langle F\rangle)/T$. Note that the
total entropy also comprises an ideal-mixing contribution of the
form $1-\ln\rho -x_A\ln x_A -x_B\ln x_B$, where $x_A$ represents
the mole fraction of species $A$ in the mixture. There is a second
way to access properties, directly from the microcanonical
ensemble. For example, the internal energy as a function of
temperature can be derived by differentiation of the entropy (i.e.
$k_B \ln \Omega(E)$) with respect to energy and and subsequent
inversion of that curve. This relies on the microcanonical
definition of temperature, namely
$T=\langle\frac{dS}{dE}\rangle^{-1}$.
\section{Results}
\begin{figure}
 \includegraphics[height=6cm]{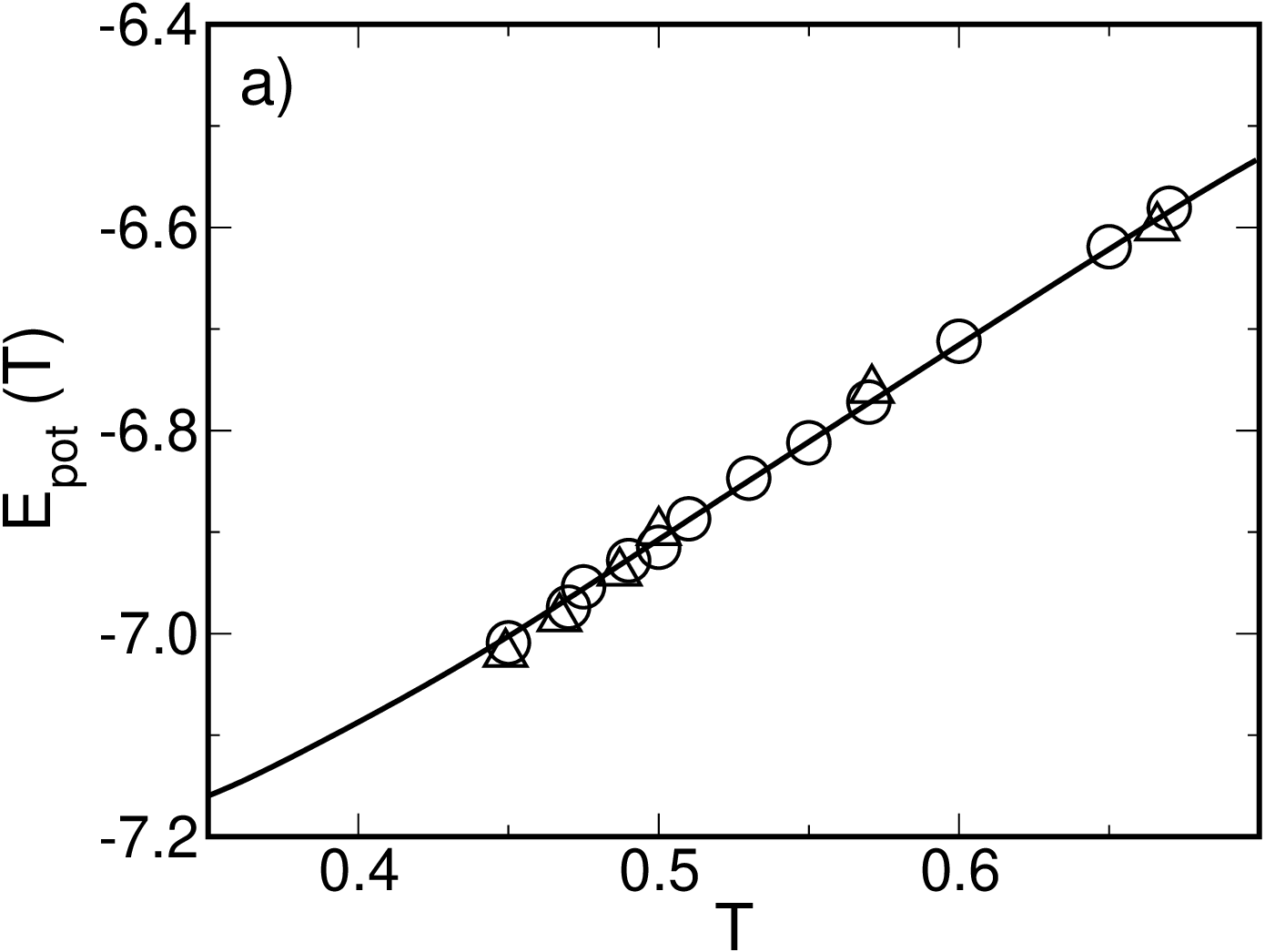}\hspace{0.5cm}
 \includegraphics[height=6cm]{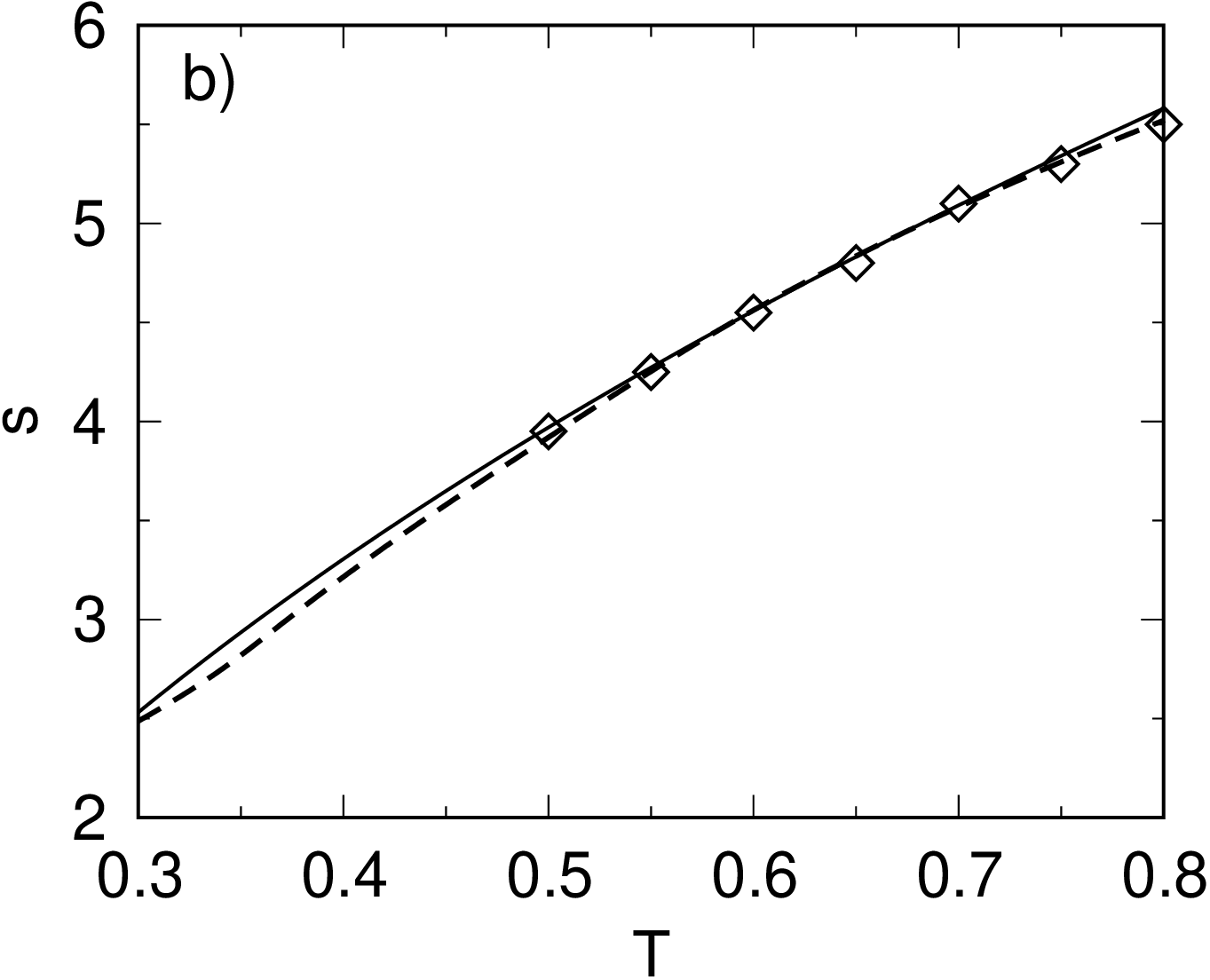}
 \caption{ a) Energy per particle calculated by canonical MC (circles), and
  density of states (solid line). Triangles are literature
  values~\cite{yamamoto00}. b) Entropy per particle. The dashed line shows the
  total entropy (excess plus ideal) determined from the density of states. The
  symbols are literature results~\cite{coluzzi00b}. The solid line is a fit to
  the literature data of the form $S=10.27*T^{0.4}-3.81$.}
 \label{fig:energy-entropy}
\end{figure}

Figure~\ref{fig:energy-entropy}a shows the average internal energy
of the binary Lennard-Jones glass former. Results by Yamamoto et
al.~\cite{yamamoto00} obtained by replica exchange Molecular
Dynamics are also shown in that figure. The agreement between the
two sets of data is quantitative. At temperatures below $T_{MCT}$
we have performed additional, extensive simulations using
biased-sampling ideas and parallel-tempering techniques. More
specifically, the algorithms developed for these additional
calculations use two-dimensional parallel tempering in
temperature~\cite{marinari92,tesi96,hansmann97,wu99,yan99} and
Hamiltonian~\cite{bunker01}, and identity swap
moves~\cite{grigera01} augmented with configurational bias
sampling~\cite{depablo92}. These techniques permit simulations at
temperatures below $T_{MCT}$, but become increasingly sluggish as
temperature is decreased. Still, the agreement for the energy
generated by those simulations and the Density of States technique
is also good, thereby providing further consistency tests for the
results presented here.

The entropy of the binary Lennard-Jones glass former is shown in
Fig.~\ref{fig:energy-entropy}b as a function of temperature. The
points in the figure represent literature data generated by
thermodynamic integration~\cite{coluzzi00b}; the entropies
simulated in this work have been shifted by a constant to make
them coincide with those data in the range $0.5<T<0.8$. The
agreement between our results and those of Coluzzi et al. is
excellent. In order to arrive at an estimate of the Kauzmann
temperature, these authors extrapolate the liquid phase entropy
below $T=0.5$ using an expression of the form $S=aT^{0.4}+b$. We
find that such a functional form is in good agreement with our
simulations; at lower temperatures, however, minor but systematic
departures from our results are observed. This would be indicative
of a slightly higher Kauzmann temperature than that reported in
the literature ($T_K\approx0.3$)~\cite{sciortino99,coluzzi00b}, as the simulated
entropy decays more rapidly than that anticipated by
extrapolation.
\section{Discussion}
The density of states as a function of temperature does not show
any unexpected behavior over the entire energy region considered
in this work. Its logarithm simply becomes steeper with decreasing
energy (see Figure \ref{fig:dos}b), reflecting the fact that the
number of accessible states becomes smaller. The system could
conceivably undergo a gas-liquid (or gas-glass) phase transition
at very low temperatures. To address this point we have also
determined the pressure. Our results suggest that such a
transition can occur at $T\approx 0.2$, where the pressure of the
glass becomes equal to that of the gas $p \approx 0$. For the same
system (but without cutoff corrections), Coluzzi and Parisi
estimated such a transition at $T\approx 0.3$. The occurrence of a
demixing transition can be ruled out by the shape of the various
pair distribution functions (not shown here), which remain
qualitatively unchanged in the range $0.2<T<0.6$. A very similar
system has been found to crystallize at such low
temperatures~\cite{middleton01}. However, crystallization is
avoided in our calculations by restricting the simulations to
energy ranges above those of the crystal. Moreover,
crystallization has only been found in constant pressure
simulations. The volume is kept constant in this work, thereby
leading to frustration of crystallization.

The results presented in this work suggest that a Monte Carlo
technique based on the concept of entropic sampling is capable of
generating high-accuracy estimates of the equilibrium density of
states of a binary Lennard-Jones glass former, down to
temperatures below the mode coupling temperature, a region of
temperature that previous studies of glass-forming liquids have
avoided. With further refinement of the algorithm discussed in
this work~\cite{yan03}, we expect that reliable simulations in the
near vicinity of the reported Kauzmann temperature will become
possible.
\section*{Acknowledgments}
RF wants to thank the Emmy-Noether Program of the Deutsche
Forschungs-Gemeinschaft for financial support.

\bibliographystyle{apsrev} \bibliography{standard}

\end{document}